# The early phase of the COVID-19 outbreak in Lombardy, Italy

Cereda D[1]*, Tirani M[1,2]*°, Rovida F[3]*, Demicheli V[4], Ajelli M[5], Poletti P[5], Trentini F[5], Guzzetta G[5], Marziano V[5], Barone A[6], Magoni M[7], Deandrea S[2], Diurno G[1], Lombardo M[8], Faccini M[4], Pan A[9], Bruno R[10,11], Pariani E[12], Grasselli G[13,14], Piatti A[1], Gramegna M[1], Baldanti F[3,11]#, Melegaro A[15,16]#, Merler S[5]#

* These authors contributed equally to this work
# These authors are joint senior authors

1 Directorate General for Health, Lombardy Region, Milano, Italy

2 Department of Hygiene and Preventive Medicine, Health Protection Agency of Pavia, Pavia, Italy

3 Molecular Virology Unit, Fondazione IRCCS Policlinico San Matteo, Pavia, Italy

4 Health Protection Agency of the Metropolitan Area of Milan, Milano, Italy

5 Center for Information and Communication Technology, Bruno Kessler Foundation, Trento, Italy

6 Regional Agency for Innovation and Procurement, Milano, Italy

7 Epidemiology Unit, Health Protection Agency of Brescia, Brescia, Italy

8 ASST Lodi, Lodi, Italy

9 Infectious Diseases Unit, ASST di Cremona, Cremona, Italy

10 Infectious Diseases Unit, Fondazione IRCCS Policlinico San Matteo, Italy

11 Department of Clinical, Surgical, Diagnostics and Pediatric Sciences, University of Pavia, Pavia, Italy

12 Department of Biomedical Sciences for Health, University of Milan, Milano, Italy

13 Department of Pathophysiology and Transplantation, University of Milan, Milano, Italy

14 Department of Anesthesia, Intensive Care and Emergency, Fondazione IRCCS Ca' Granda Ospedale Maggiore Policlinico, Milano, Italy

15 Department of Social and Political Sciences, Bocconi University, Milano, Italy

16 Carlo F. Dondena Centre for Research on Social Dynamics and Public Policy, Bocconi University, Milano, Italy

°Corrisponding author:

Marcello Tirani

Directorate General for Health, Lombardy Region,

Piazza Città di Lombardia 1, 20124, Milano

+39 333.3553594

marcello_tirani@regione.lombardia.it

# Abstract

**Background**

In the night of February 20, 2020, the first case of novel coronavirus disease (COVID-19) was confirmed in the Lombardy Region, Italy. In the week that followed, Lombardy experienced a very rapid increase in the number of cases. We analyzed the first 5,830 laboratory-confirmed cases to provide the first epidemiological characterization of a COVID-19 outbreak in a Western Country.

**Methods**

Epidemiological data were collected through standardized interviews of confirmed cases and their close contacts. We collected demographic backgrounds, dates of symptom onset, clinical features, respiratory tract specimen results, hospitalization, contact tracing. We provide estimates of the reproduction number and serial interval.

**Results**

The epidemic in Italy started much earlier than February 20, 2020. At the time of detection of the first COVID-19 case, the epidemic had already spread in most municipalities of Southern-Lombardy. The median age for of cases is 69 years (range, 1 month to 101 years). 47% of positive subjects were hospitalized. Among these, 18% required intensive care. The mean serial interval is estimated to be 6.6 days (95% CI, 0.7 to 19). We estimate the basic reproduction number at 3.1 (95% CI, 2.9 to 3.2). We estimated a decreasing trend in the net reproduction number starting around February 20, 2020. We did not observe significantly different viral loads in nasal swabs between symptomatic and asymptomatic.

**Conclusions**

The transmission potential of COVID-19 is very high and the number of critical cases may become largely unsustainable for the healthcare system in a very short-time horizon. We observed a slight decrease of the reproduction number, possibly connected with an increased population awareness and early effect of interventions. Aggressive containment strategies are required to control COVID-19 spread and catastrophic outcomes for the healthcare system.

# Introduction

In the late night of February 20, 2020, the first case of novel coronavirus infectious disease (CoVID-19) was confirmed in the Lombardy Region, Italy, a district with 10 million inhabitants. The patient was a 38-year old healthy man (*patient 1*), admitted to the Hospital of Codogno, with a mild pneumonia resistant to therapy, no relevant travel history and no apparent exposure to diseased contacts. Starting with the morning of February 21, a strict proactive intervention for tracing and testing all patient' contacts was set up. Although the *patient zero* could not be detected, an initial outbreak was identified around the city of Codogno. In the week that followed, Codogno area, as well as several neighboring towns in southern Lombardy, experienced a very rapid increase in the number of detected cases, which rose to over 530 positive samples by February 28 and 5,830 by March 8$^{th}$. With the increase in the number of detected cases, also their spatial distribution expanded, reaching several other areas in northern Lombardy. Over this period, the Regional Health System, Local Health Authorities (*ATS: Agenzia di Tutela della Salute*), the Lombardy Reference Virology laboratories, and the National Health Institute (*ISS: Istituto Superiore della Sanità*) collaborated to introduce coordinated actions to limit the spread of the infection. Among these, isolation of cases, contact tracing, intensive testing, and the definition of a "red" quarantined area around the most affected towns.

Here we provide an analysis of the first 5,830 laboratory-confirmed cases reported in Lombardy, with date of symptoms onset over the period from January 14 to March 8, 2020. Epidemiological analyses of the confirmed cases and their background demographic and exposure characteristics are presented here as well as the transmission dynamics of the infection within the Region. Also, the virological analysis on a subsample of the reported cases is included to provide preliminary assessment of the level of the viral load among symptomatic and asymptomatic cases.

# Method

The earliest case was detected through the Lombardy Notifiable Disease Surveillance System, which is routinely used to monitor about 80 infectious diseases including at the time of reporting COVID-19. Due to the relevance of this diagnosis for immediate intervention, the data was simultaneously confirmed by real time RT PCR by two independent Regional Reference Laboratories, and further confirmed by sequencing of the amplicons. As indicated by the ISS, the sample was then sent to the ISS Rome Laboratory for confirmatory testing. Nevertheless, immediate intervention measures were set up on the basis of the local laboratory data.

*Case definition*

The case definitions for suspected COVID-19 cases were based on the European Centre for Disease Prevention and Control (ECDC) case definitions. A suspected COVID-19 case was defined according to the following criteria:

1) a patient with acute respiratory tract infection (sudden onset of at least one of the following: cough, fever, shortness of breath) AND with no other aetiology that fully explains the clinical presentation AND at least one of these other conditions: a history of travel to or residence in China, OR patient is a health care worker who has been working in an environment where severe acute respiratory infections of unknown etiology are being cared for; OR

2) a patient with any acute respiratory illness AND at least one of these other conditions: having been in close contact with a confirmed or probable COVID-19 case in the last 14 days prior to onset of symptoms, OR having visited or worked in a live animal market in Wuhan, Hubei Province, China in the last 14 days prior to onset of symptoms, OR having worked or attended a health care facility in the last 14 days prior to onset of symptoms where patients with hospital-associated COVID-19 have been reported.

A probable case was defined as a suspected case for whom testing for virus causing COVID-19 is inconclusive (according to the test results reported by the laboratory) or for whom testing was positive with a specific real-time RT PCR assay detecting the SARS-CoV-2 virus responsible for the COVID-19. Confirmed case was a person with laboratory confirmation of virus causing of SARS-CoV-2 infection, irrespective of clinical signs and symptoms.

*Epidemiological data and statistical analysis*

Epidemiological data were collected through standardized interviews of confirmed cases and their close contacts performed by the ATS operators. The information gathered were used to fill the datasets and included: demographic backgrounds, dates of symptom onset, clinical features, respiratory tract specimen results, hospitalization, contact tracing. Investigators interviewed each positive subject and/or their relatives, where necessary, to determine the history of exposure during the 2 weeks before the symptom onset of the positive test result, including dates, times, relationships, hobbies, and other patterns of exposures. The definition of contact used in the epidemiological field investigation is provided in the *Supplementary Material*. Data from the local datasets were entered into a central database and analyzed. Cases with established epidemiological links were used to estimate the distribution of the serial interval (i.e., the time period between the time of onset of symptoms in a primary case and the time of onset in her/his secondary cases.

*Laboratory testing*

According to WHO suggestions, nasal swabs (UTM viral transport ®, Copan Italia S.p.a) from all suspected cases were tested with at least two real-time RT PCR assays targeting different genes (E and RdRp) of SARS-CoV-2 [1]. In addition, a novel quantitative real-time RT PCR targeting an additional SARS-CoV-2 gene (M) was developed (details provided upon request). From February 21 to February 25, all suspected cases and asymptomatic contacts were tested. From February 26 onward, testing was applied only to symptomatic patients.

*COVID-19 transmission dynamics*

The basic reproduction number $R_0$ represents the average number of secondary cases generated by a primary infector in a fully susceptible population. Once interventions are put in place or the number of susceptible individuals declines, the transmission potential of the disease at a given time t is measured in terms of the net reproduction number *Rt*. We provide estimates of both *$R_0$* and *Rt* for the entire Lombardy and for the three most affected areas by using a Bayesian approach widely adopted in the literature [2-4] (details reported in the *Supplementary Material*).

**Ethical approval**

Data collection and analysis of cases and close contacts was part of a continuing public health outbreak investigation and were thus considered exempt from institutional review board approval.

**Results**

As of March 8, 2020, a total of 5,830 positive cases were reported in all provinces of the Lombardy Region (Figure 1). The first case (*patient 1*) was detected and laboratory confirmed at the Hospital of Codogno in the Lodi Province on February 20. By the following day 28 positive cases were identified in the same area. On February 23, given the upsurge of positive cases in the area, a Prime Minister Decree (DPCM Law 23/02/20 n.6) introduced strict measures aimed at containing the spread of the disease. The implemented measures included the lockdown of a number of municipalities that appeared at the center of the outbreak (Red Zone), the closure of schools of all grades and universities, and the suspension of public activities, sports events, and social gatherings across the Region.

The rapid intensification of regional surveillance that occurred in the following days, through contact tracing and testing of both symptomatic and asymptomatic exposures to positive cases, provided critical information for the detection of possible epidemiological links, and uncovered ongoing transmission previous to the identification of *patient 1*. For this analysis, it emerged a highly different figure of the Lombardy outbreak with ongoing transmission in January and much less steep in the number of cases as it could misleadingly be suggested by the temporal trend in the number of notified cases (see *Supplementary Material*). A delay of 3.6 days (95% CI, 1 to 10) was found between the date in which the result of the test was received and the date of the recording in the dataset. The delay between the dates of symptom onset and reporting was 7.3 days (95%CI, 1 to 20). Therefore, the dates of symptom onset provide a reliable picture of the situation up to about the end of February. Both delays showed a decline over time (see *Supplementary Material*).

During the early stages of the COVID-19 epidemic in Lombardy, we observed the formation of three major clusters identified around the cities of Codogno, Bergamo, and Cremona (see *Supplementary Material*). Later on, the epidemic started to widespread in the entire region (Fig. 2). However, as of March 5, the majority of cases (72%) were observed in the provinces of Bergamo, Lodi (where Codogno is located), and Cremona.

We analyzed the characteristics of the confirmed cases stratified by three time periods: the first period accounts for patients with symptom onset up to February 19, the day before the first case was identified in Codogno; the second for those with symptom onset between February 20 and February 25, the day before a change in the testing policy was introduced entailing the stop of testing for asymptomatic contacts of COVID-19 cases; the third period considers cases with symptom onset from February 26 onwards. For consistency with the rest of the analysis, we excluded those cases that were explicitly reported as *asymptomatic* (n=204) as this group was more affected by changes in the testing and reporting criteria over the considered periods. From the remaining pool of records (n=5,626), we found a median age for positive cases of 69 years (range, 1 month to 101 years), with over half of the cases occurring in the over 65 (34% in the 75+) and 62% of the overall number of positives being males (Table 1). Almost half of the cases were hospitalized (47%), although this percentage increases when we stratify the dataset considering those reporting a date of symptom onset and those who did not (most likely less severe cases). In the former group 63%, 61% and 56% of the cases were hospitalized in, respectively, the three considered periods whereas in the latter group (those without a date of symptom onset) the percentage was 29%. Among those hospitalized, 18% required intensive care. Overall, 346 deaths occurred in the region and the case fatality rates (CFRs) was highest in the older age groups (14% in the 75+ years). Over the periods we observed changes in the geographical distribution of the cases in the twelve Provinces of the Lombardy Region, with Lodi being the most affected province in Period 1 and the fourth in Period 3, after Bergamo, Brescia and Cremona which, respectively, accounted for 25%, 21% and 17% of the overall number of cases of the region.

From the analysis of 90 observations of individual serial intervals in 55 clusters, we estimated the distribution of the serial interval to follow a gamma distribution with mean 6.6 days (percentiles $2.5^{th}$ and $97.5^{th}$ of the distribution: 0.7-19.0). Hence, we estimated that 95% of cases develop symptoms within 16.1 days of their infector.

For the entire Lombardy and in each of the three analyzed major clusters, we identified an initial phase lasting between one and two weeks characterized by exponential growth. During that phase, we estimated the doubling time of the epidemic to be 3.1 days (95% CI, 2.2 to 5.4) in Bergamo, 3.5 days (95% CI, 2.7 to 4.8) in Codogno, and 3.4 days (95% CI, 2.5 to 5.0) in Cremona. In Lombardy as a whole, the epidemic doubling time was estimated ad 2.6 days (95% CI, 2.2 to 3.0). In the same phase, we estimated the basic reproduction number at 2.9 (95% CI, 2.3 to 3.4) in Bergamo, 2.5 (95% CI, 2.1 to 2.8) in Codogno, 2.3 (95% CI, 1.9 to 2.7) in Cremona, and 3.1 (95% CI, 2.9 to 3.2) in Lombardy as a whole. In all locations, the exponential growth phase of the epidemic did not last long and the net reproduction number $Rt$ was estimated to follow a highly variable temporal dynamic. In particular, in Bergamo and Cremona, we estimated a rapid increase of $Rt$ before starting to slightly decrease (Fig. 3). In Codogno, the first reported cases developed symptoms in late January, 2020. Since then, *Rt* was estimated to slowly, but constantly, increase for more than 2 weeks before starting to decline (Fig. 3). In Lombardy as a whole, we estimated *Rt* to be under the epidemic threshold for the initial part of the epidemic and then to quickly grow as soon as the epidemic started to spread in different geographical areas (Fig. 3). Most importantly, in all locations we estimated *Rt* to follow a decreasing trend since February 20, although it is estimated to be still above the epidemic threshold. These results are robust to assuming different distributions of the serial interval (see *Supplementary Material*).

Overall, from February 21 to March 3, 7,925 nasal swabs were analyzed in Lombardy, with 2,251 positives for SARS-Cov-2. A deeper analysis was performed on the subset data from Pavia Reference Laboratory where 6,208 subjects were swabbed and 1,529 resulted positive. Out of 1,529 positive subjects, 422 (27.6%) were tested between February 21 and February 26, when both symptomatic patients and symptomatic/asymptomatic contacts were considered (the positivity rate was 14.6%, 422/2,892). In particular, among the 380 positive subjects detected on February 21, 17 (4.5%) were defined as asymptomatic, 295 (77.6%) were classified as symptomatic, and in 68 cases (17.8%) there was no clear indication on the presence of symptoms. Over the period from February 27 to March 3, only symptomatic primary cases and their symptomatic contacts were tested, increasing the detection rate to 33.4% (1,107/3,316).

The median SARS-CoV-2 levels in nasal swabs was 5.0 Log10 RNA copies/ml (range 1.7-10.1) in symptomatic and 4.7 log10 RNA copies/ml (range 2.1-7.1) in asymptomatic patients (Fig. 4). The viral load in nasal swabs of symptomatic patients and asymptomatic subjects was not statistically different (two-sample t-test, p=0.51).

**Discussion**

We have provided an initial assessment of the rapidly changing epidemiology and transmission dynamics of COVID-19 outbreak occurred in Lombardy Region. What we have observed is an impressive increase in positive cases within a few days, of which about half needed hospitalization and 16% of these required admission to intensive care unit.

At the beginning of the epidemic, cases were mostly observed in Codogno, within the Lodi Province, in the southern part of Lombardy. As the epidemic progressed, the infection spread to other areas of the region, causing significant and increasing burden of disease and intensive care needs, especially among the elderly. This confirms findings on COVID-19 epidemiology in China [5]. However, with the data available at this time, it is not possible to assess whether the lack of cases among children and young adults is a consequence of reduced risk of infection or a propensity for milder clinical symptoms. Moreover, we found that about 60% of COVID-19 cases were male, although the reason remains to be clarified.

We estimated the mean serial interval to be 6.6 days, which lies in between the estimates obtained for Hubei Province of China (mean 7.5 days) [6] and for the other Chinese provinces (mean 5.1 days) [4]. The observed difference of 1.5 days with respect of the mean serial interval in Chinese provinces outside Hubei is a possible indicator that COVID-19 cases in Lombardy are not isolated as quickly as in those provinces of China and thus continue to spread the infection for a longer amount of time. We estimated the mean $R_0$ to be in the range 2.3-3.1 during the exponential growth phase of the epidemic. The daily reproduction number has highly fluctuated over time, but it is rather homogenous across geographical areas. In all the three identified clusters and in Lombardy as a whole, we estimate a slightly decreasing trend of $Rt$ starting form about February 20. However, we estimated $Rt$ to be still well above the epidemic threshold.

We did not observe significantly different viral loads in nasal swabs between symptomatic and asymptomatic subjects, suggesting the same potential for transmitting the virus. However, the limited number of asymptomatic infected subjects that were identified through contact tracing during the first week of data collection may suggest a minor role of asymptomatic individuals in overall spread of the infection. In contrast, active search of both symptomatic and asymptomatic cases in the initial phases of the outbreak drained significant resources somewhat limiting the timely identification of all infected individuals.

It is important to note that this study is clearly affected by the usual limitations deriving from the data analysis of rapidly evolving infectious disease outbreaks. In such a situation the proper identification of the correct epidemiological links between cases is challenging. This increases the uncertainty surrounding our estimates of the serial interval. As time progressed, critical public health interventions were put in place affecting the quickness of data collection. This partially explains the large delay between the dates of symptom onset and reporting. Such an observed delay does not allow the analysis of the net reproduction number to be extended after February 25 (as the data after that date could be incomplete). The estimates of the net reproduction number are robust to an unknown level of underreporting, but would be affected by a time-varying reporting rate. This could be the case here, given the change in the testing protocol occurred on February 25. Asymptomatic transmission remains a challenge to prove and quantify, although we found virological evidence suggesting a similar viral load in symptomatic and asymptomatic subjects. This may also affect the estimates of the reproduction number.

The threat posed by COVID-19 is unprecedented and greatly complicated by the limited knowledge on COVID-19 epidemiology. Policy makers in the Lombardy Region had to take rapid decisions on how to mitigate the spread of the epidemic, minimizing morbidity and mortality, delaying an epidemic peak that was going to overwhelm healthcare services. The initial response of the Regional Health System was almost immediate and was articulated around three main objectives: collection of epidemiological data to understand the situation and perform model-based predictions, increase in diagnostics (strengthening of regional laboratories), and promoting hospital assistance for affected subjects (suspension of outpatient activities, creation of dedicated paths and areas, upgrading of hospital and intensive care places). Coordinated actions were taken to limit the spread of the infection (e.g., case isolation, contact tracing, and the definition of a quarantined area). Despite these interventions, we have observed a progressive and parallel growth in the incidence, hospitalizations, hospital and intensive care efforts, and mortality. All of those are responsible for a rapid saturation of the health emergency system with a progressive difficulty of being able to treat both subjects with COVID-19 and those with other pathologies. Nonetheless, as shown by our analysis, the set-up of a quarantine area around the epicenter of the outbreak in Codogno, appears to have played a critical role in controlling the infection, inducing a reduction in the number of reported positive cases in the area and a consequent decrease in the estimate of the net reproductive number. This appeared as relatively robust evidence that helped informing the decision of enforcing other stringent measures to the entire Region. We expect that the impact of these measures may be more markedly visible in the next couple of weeks.

In conclusion, even in the presence of all considered limitations, this analysis of the first two weeks of data provides important insights into the transmission dynamics of the infection, the virological characteristics of positive cases, both asymptomatic and symptomatic, as well as the critical factors in the reporting system that need to be further considered. Such information will be highly valuable for other Italian regions and countries that are now facing an upsurge in the number of COVID-19 cases.

**References**

1. Corman VM, Landt O, Kaiser M, Molenkamp R, Meijer A, Chu DKW et al. Detection of 2019 novel coronavirus (2019-nCoV) by real-time RT-PCR. Eurosurveill. 2020; 25(3). doi: 10.2807/1560-7917.ES.2020.25.3.2000045.

2. World Health Organization Ebola Response Team, Ebola virus disease in West Africa--the first 9 months of the epidemic and forward projections. N Engl J Med 2014; 371(16): 1481-95.

3. Liu Q-H, et al. Measurability of the epidemic reproduction number in data-driven contact networks. Proc Natl Acad Sci USA 2018; 115(50): 12680.

4. Zhang J, et al. Evolving epidemiology of novel coronavirus diseases 2019 and possible interruption of local transmission outside Hubei Province in China: a descriptive and modeling study. medRxiv 2020; doi: https://doi.org/10.1101/2020.02.21.20026328

5. Chen N, Zhou M, Dong X, et al. Epidemiological and clinical characteristics of 99 cases of 2019 novel coronavirus pneumonia in Wuhan, China: a descriptive study. Lancet 2020.

6. Li Q, et al. Early transmission dynamics in Wuhan, China, of novel coronavirus–infected pneumonia. N Engl J Med 2020.

**Table 1** - Characteristics of patients with COVID-19 in Lombardy Region, Italy, as of March 8, 2020*

| | | Period of the epidemic | | | |
|---|---|---|---|---|---|
| | All (n=5626) | Period 1 Until Feb19 (n=388) | Period 2 Feb 20 - Feb25 (n=1135) | Period 3 Feb26 - Mar5 (n=1998) | Uncategorized (n=2105) |
| **Demographics** | | | | | |
| Median age - yr (range) | 69 (0-101) | 71 (1-99) | 69 (0-97) | 72 (0-101) | 67 (0-101) |
| Male sex - no./total no. (%) | 3510/5626 (62) | 266/388 (69) | 761/1135 (67) | 1239/1998 (62) | 1244/2105 (59) |
| Health care workers - no./total no. (%) | 888/5,626 (16) | 22/388 (6) | 86/1,135 (8) | 174/1,998 (9) | 306/2,105 (15) |
| Age group - no./total no. (%) | | | | | |
| <18 yr | 44/5529 (1) | 3/388 (1) | 7/1135 (1) | 17/1997 (1) | 17/2009 (1) |
| 18-24 yr | 50/5529 (1) | 2/388 (1) | 12/1135 (1) | 10/1997 (1) | 26/2009 (1) |
| 25-49 yr | 914/5529 (17) | 56/388 (14) | 187/1135 (16) | 277/1997 (14) | 394/2009 (20) |
| 50-64 yr | 1464/5529 (26) | 100/388 (26) | 288/1135 (25) | 485/1997 (24) | 591/2009 (29) |
| 65-74 yr | 1185/5529 (21) | 98/388 (25) | 296/1135 (26) | 421/1997 (21) | 370/2009 (18) |
| 75+ yr | 1872/5529 (34) | 129/388 (33) | 345/1135 (30) | 787/1997 (39) | 611/2009 (30) |
| Hospitalised no./total no. (%) | 2,669/5,626 (47) | 245/388 (63) | 692/1,135 (61) | 1,113/1,998 (56) | 619/2,105 (29) |
| ICU no./total no. in H (%) | 440/2669 (16) | 30/388 (12) | 119/692 (17) | 190/1,113 (17) | 101/619 (16) |
| Fatalities - no./total no. (%) | 343/5626 (6) | 41/388 (10) | 105/1,135 (9) | 131/1,998 (6) | 66/2,105 (3) |
| <55 yr | 0/1296 (0) | 0/65 (0) | 0/276 (0) | 0/403 (0) | 0/552 (0) |
| 55-64 yr | 9/1,179 (1) | 1/89 (1) | 0/203 (0) | 3/371 (1) | 5/516 (1) |
| 65-74 yr | 63/1,197 (5) | 9/95 (9) | 23/301 (8) | 19/403 (5) | 12/398 (3) |
| 75+ yr | 271/1,954 (14) | 31/139 (22) | 82/355 (8) | 109/821 (5) | 49/639 (8) |
| **Provinces of the Lombardy Region** | | | | | |
| Bergamo | 1231/4805 (26) | 91/385 (24) | 307/1106 (28) | 462/1865 (25) | 371/1449 (26) |
| Cremona | 910/4805 (19) | 59/385 (15) | 189/1106 (17) | 325/1865 (17) | 337/1449 (23) |
| Lodi | 801/4805 (17) | 132/385 (34) | 258/1106 (23) | 212/1865 (11) | 199/1449 (14) |
| Brescia | 756/4805 (16) | 38/385 (10) | 120/1106 (11) | 396/1865 (21) | 202/1449 (14) |
| Milano | 506/4805 (11) | 30/385 (8) | 110/1106 (10) | 197/1865 (11) | 169/1449 (12) |
| Pavia | 287/4805 (6) | 21/385 (5) | 57/1106 (5) | 98/1865 (5) | 111/1449 (8) |
| Mantova | 98/4805 (2) | 4/385 (1) | 16/1106 (1) | 41/1865 (2) | 37/1449 (3) |
| Lecco | 66/4805 (1) | /385 (0) | 8/1106 (1) | 52/1865 (3) | 6/1449 (0) |
| Monza Brianza | 62/4805 (1) | 6/385 (2) | 16/1106 (1) | 34/1865 (2) | 6/1449 (0) |
| Varese | 43/4805 (1) | /385 (0) | 11/1106 (1) | 27/1865 (1) | 5/1449 (0) |
| Como | 38/4805 (1) | 3/385 (1) | 10/1106 (1) | 20/1865 (1) | 5/1449 (0) |
| Sondrio | 7/4805 (0) | 1/385 (0) | 4/1106 (0) | 1/1865 (0) | 1/1449 (0) |

*Reduced denominators indicate missing data. Percentages may not total 100 because of rounding.

**Figure captions**

**Figure 1.** Daily number of new confirmed symptomatic COVID-19 cases in Lombardy by date of onset (N=3,521). The epidemic curve is based on data up to March 8, 2020. The decline in the number of cases in the last few days is partially due to the delay between the date of reporting and the date of symptom onset. The colors of the bars represent the three considered periods.

**Figure 2.** Geographical distribution of the cumulative number of cases (dates of symptom onset) across all municipalities of Lombardy.

**Figure 3.** Estimated net reproduction number $R_t$ in the three main clusters and in Lombardy as a whole, over a 4-day moving average. We excluded cases with symptom onset after February 25, 2020 to account for the incompleteness of the dataset in the last few days due to reporting delays.

**Figure 4**. SARS-CoV-2 viral load in tested subjects classified by symptom type.

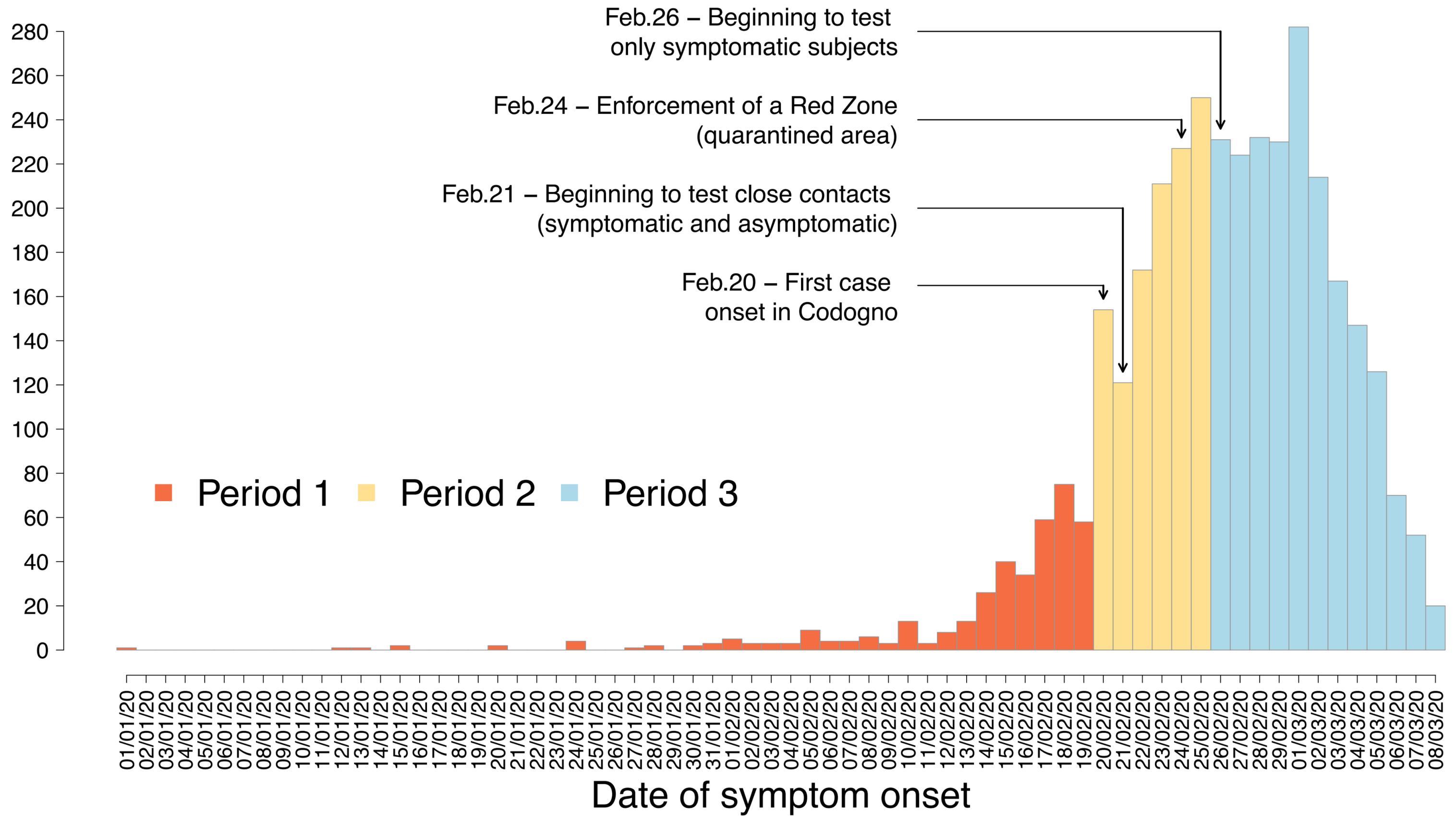

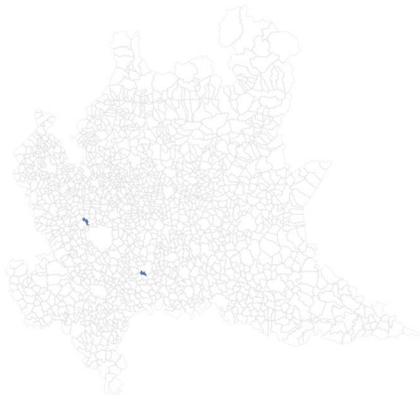 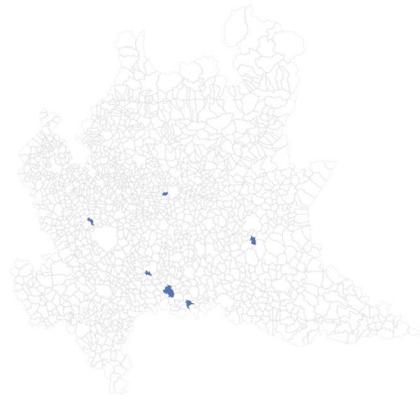 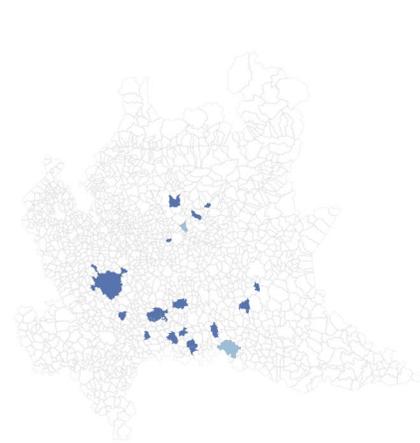 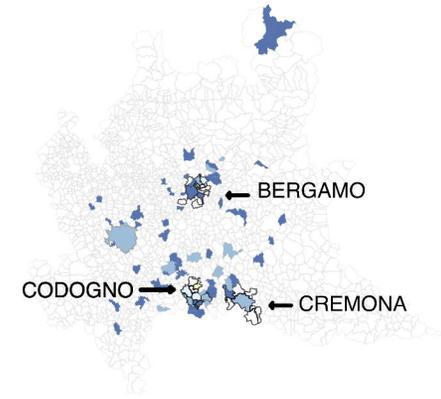
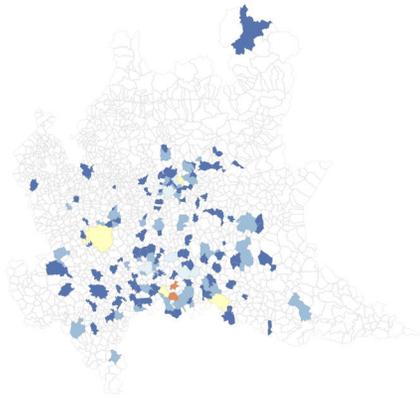 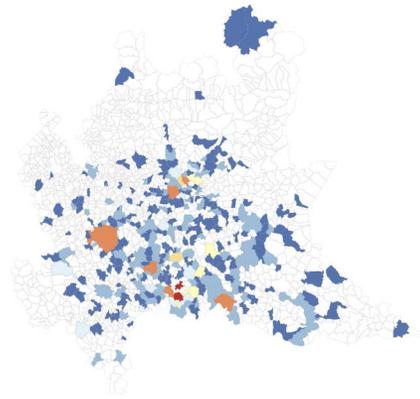 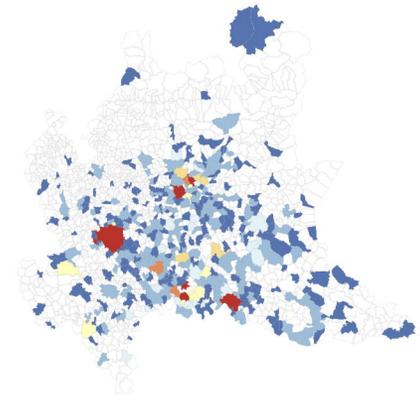 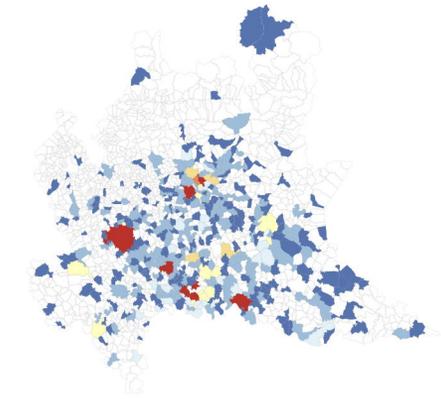
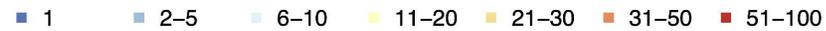

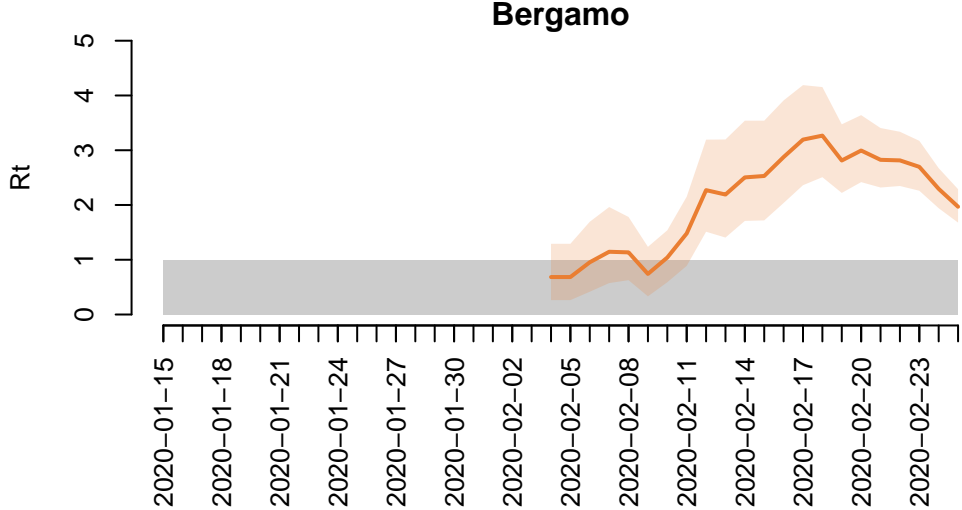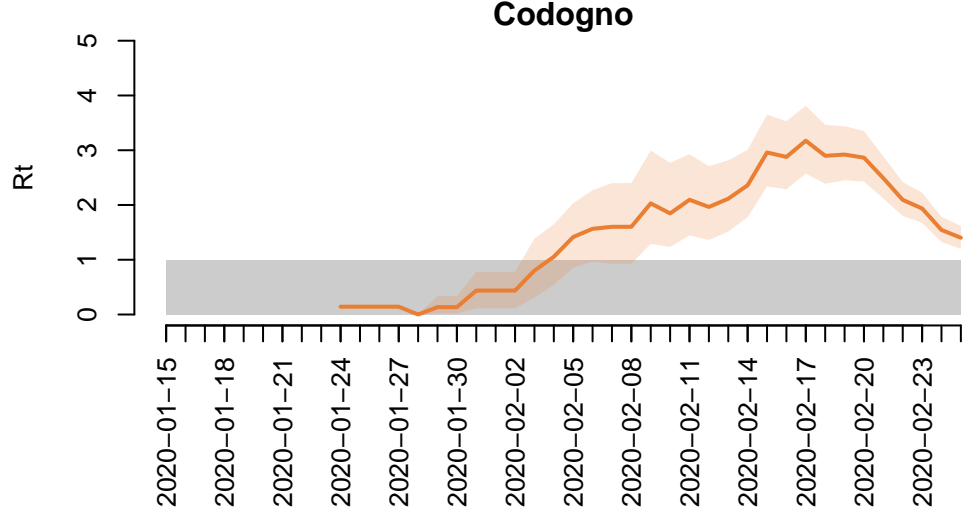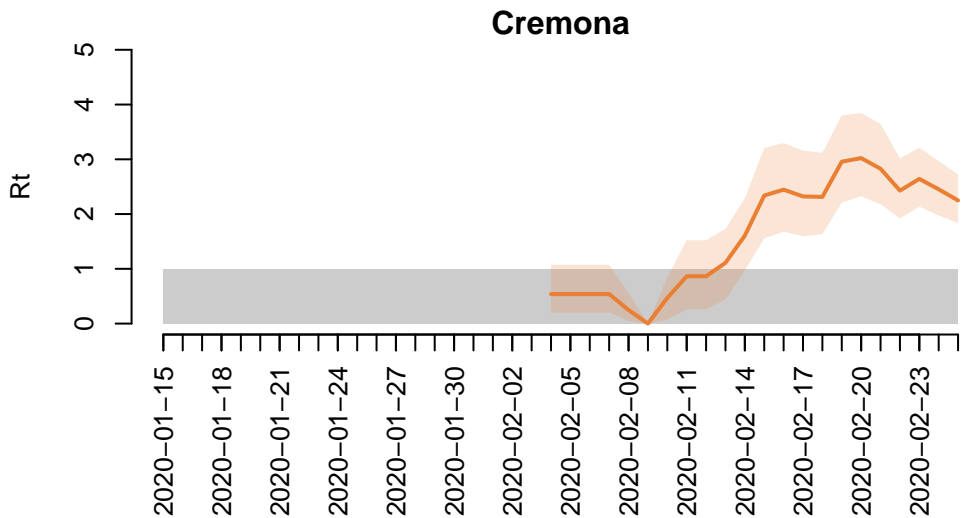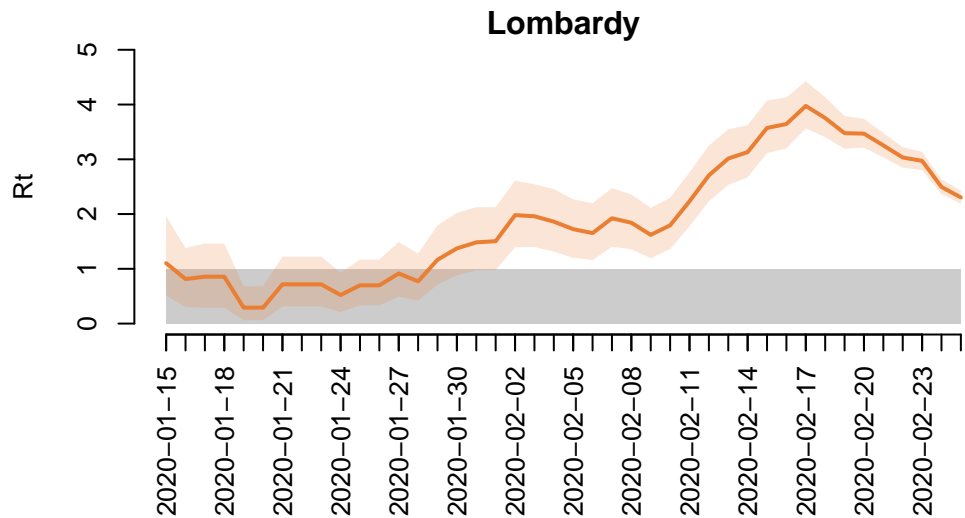

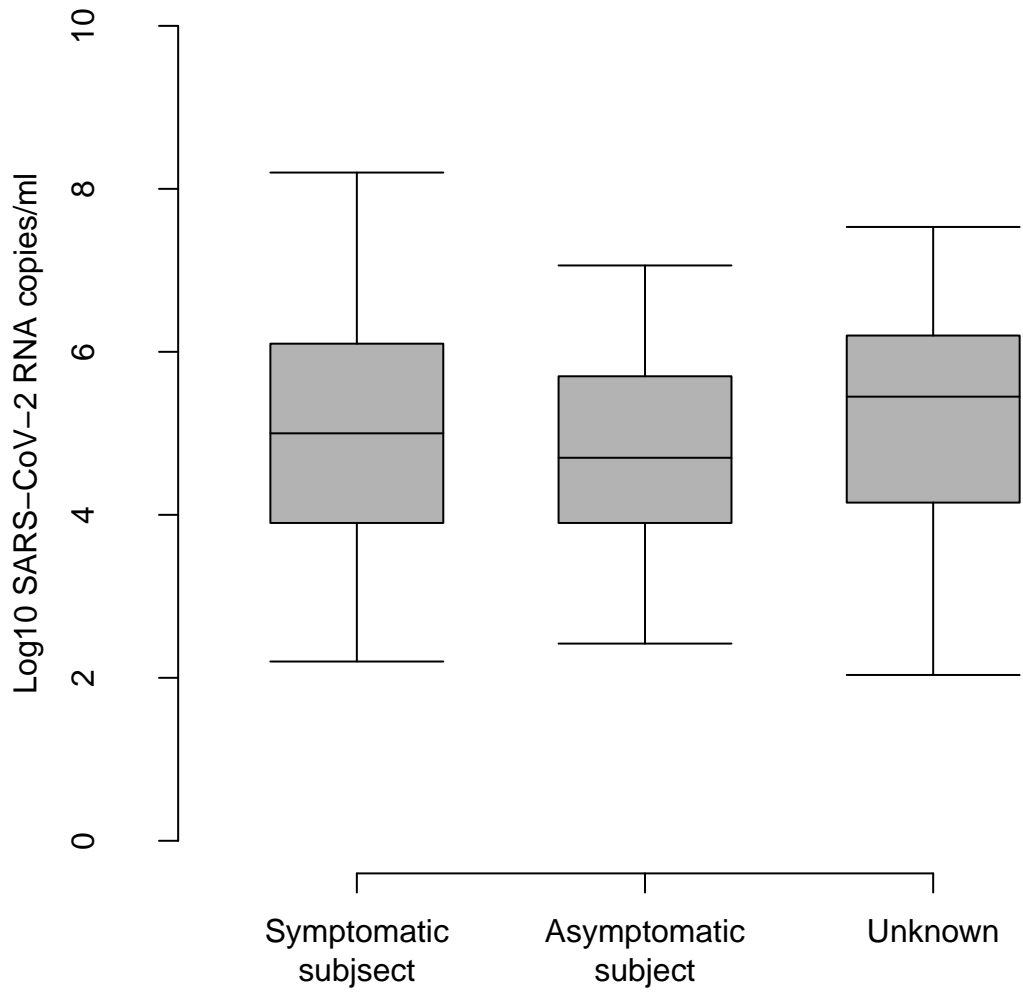

# Supplementary material

# The early phase of the COVID-19 outbreak in Lombardy, Italy

Cereda D, Tirani M, Rovida F, Demicheli V, Ajelli M, Poletti P, Trentini F, Guzzetta G, Marziano V, Barone A, Magoni M, Deandrea S, Diurno G, Lombardo M, Faccini M, Pan A, Bruno R, Pariani E, Grasselli G, Piatti A, Gramegna M, Baldanti F, Melegaro A, Merler S

## Table of Contents

## 1. Definition of contact

Close contact of a confirmed case was considered a person living in the same household as a COVID-19 case, a person having had face-to-face contact with a COVID-19 case within 2 meters and >15 minutes; a person who was in a closed environment (e.g. classroom, meeting room, hospital waiting room, etc.) with a COVID-19 case >15 minutes and at a distance of <2 meters; a healthcare worker (HCW) or other person providing direct care for a COVID-19 case, or laboratory workers handling specimens from a COVID-19 case without recommended personal protective equipment (PPE) or with a possible breach of PPE; a contact in an aircraft sitting within two seats (in any direction) of the COVID-19 case, travel companions or persons providing care, and crew members serving in the section of the aircraft where the index case was seated (if severity of symptoms or movement of the case indicate more extensive exposure, passengers seated in the entire section or all passengers on the aircraft may be considered close contacts). The epidemiological link could have occurred within a 14-day period before the onset of symptoms in the case under consideration.

## 2. Definition of geographical clusters

Three clusters of municipalities were defined around the cities of Codogno, Cremona, and Bergamo, which represent the largest cities in the three most affected geographical areas of Lombardy as of March 8, 2020. A cluster was defined as all municipalities whose barycenter lay within a radius of 10 km from the barycenter of the three major cities. The 'haversine method' was used to calculate the distance between municipalities. Table S1 shows the list of municipalities included in the three clusters. Figures S1 shows the epidemic curves associated with the three identified clusters and the contribution of the three clusters to the epidemic trajectory of the whole region.

*Table S1.* List of municipalities and population size of the three clusters. Data taken from the official Italian records [1].

| Municipality | Population | Municipality | Population | Municipality | Population |
|---|---|---|---|---|---|
| Codogno cluster | 33,454 | Bergamo cluster | 253,511 | Cremona cluster | 90,769 |
| Codogno | 15,991 | Bergamo | 121,639 | Cremona | 72,680 |
| Casalpusterlengo | 15,293 | Seriate | 25,385 | Castelverde | 5,685 |
| Castiglione D'Adda | 4,646 | Dalmine | 23,610 | Persico Dosimo | 3,389 |
| Somaglia | 3,836 | Alzano Lombardo | 13,655 | Sesto ed Uniti | 3,218 |
| Maleo | 3,098 | Nembro | 11,526 | Stagno Lombardo | 1,541 |
| Fombio | 2,317 | Scanzorosciate | 10,011 | Bonemerse | 1,503 |
| San Fiorano | 1,839 | Sorisole | 9,139 | Spinadesco | 1,476 |
| Castelgerundo | 1,498 | Torre Boldone | 8,777 | Pieve d`Olmi | 1,277 |
| Terranuova dei Passerini | 927 | Zanica | 8,739 | | |
| | | Albano Sant'Alessandro | 8,294 | | |
| | | Villa di Serio | 6,780 | | |
| | | Ranica | 5,956 | | |

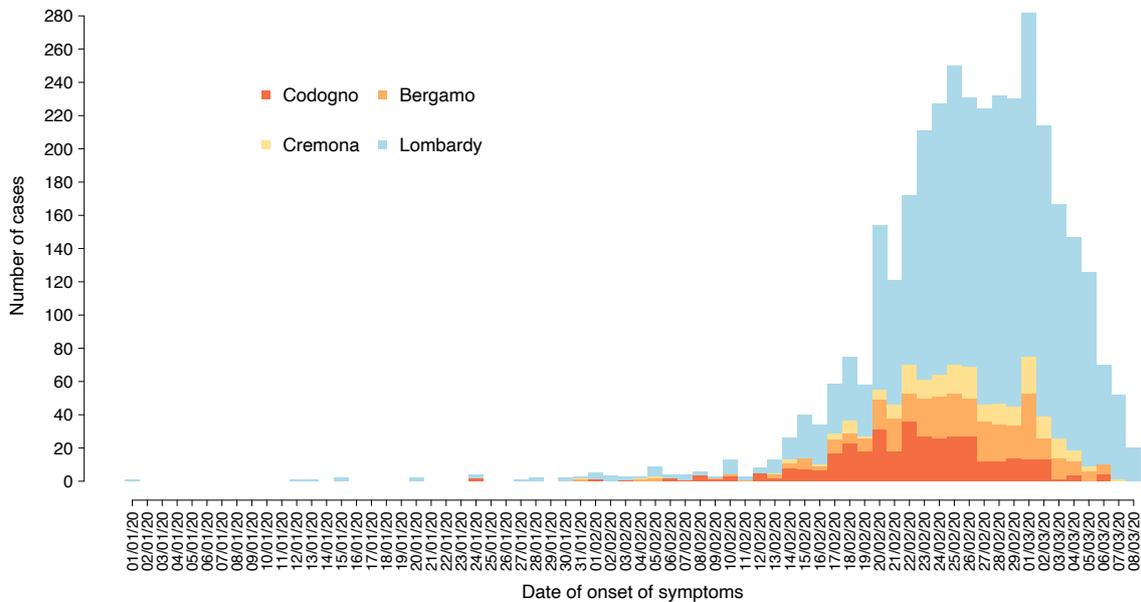

***Figure S1.*** *Daily number of positive COVID-19 cases by date of symptom onset in the three identified geographical clusters and in the remaining areas of the Lombardy.*

### 3. Reporting delays

The temporal trend of the number of reported cases provides a highly different picture of the epidemiological situation of the COVID-19 outbreak in Lombardy (Fig. S2). This is due to the delay in identification of cases and reporting. Data on laboratory-confirmed symptomatic cases reported in Lombardy by March 8, 2020 were used to estimate the time between the onset of symptoms and the laboratory confirmation. We used maximum likelihood to fit a negative binomial distribution to the data. We found that the best-fitting distribution has mean 5.25 (SD: 0.085) and dispersion 1.57 (SD: 0.054), see Figure S3.

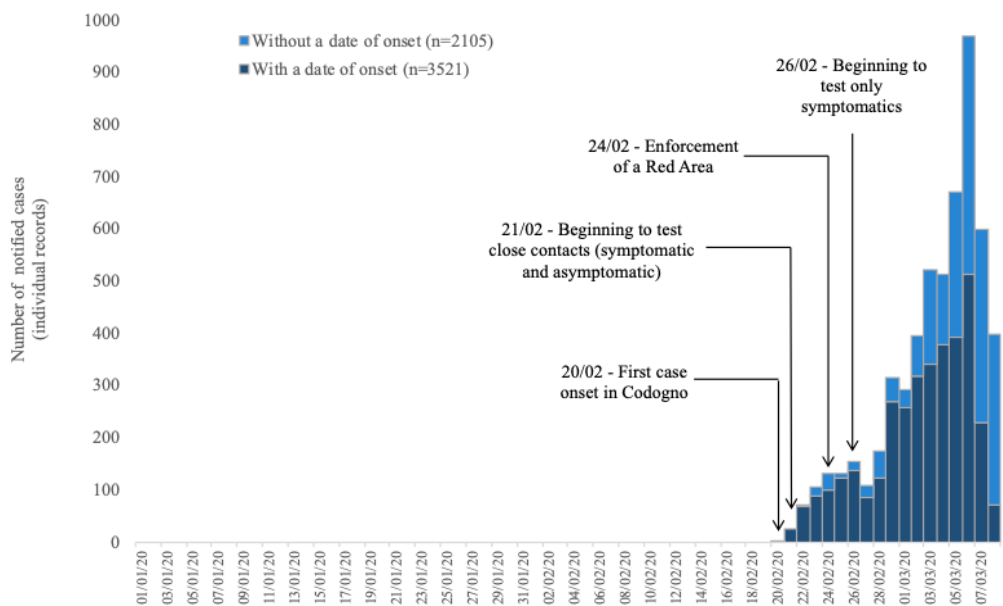

***Figure S2.*** *Confirmed symptomatic cases of COVID-2019 in Lombardy Region, Italy, by date of laboratory results (N=5,626).*

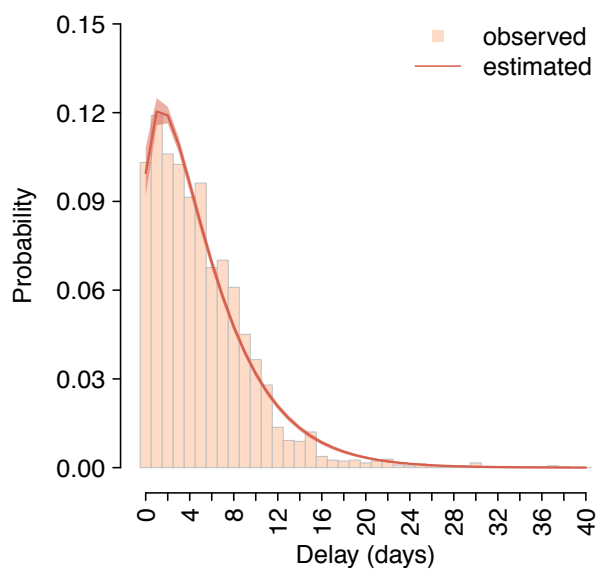

***Figure S3.*** *Distribution of the delay between the onset of symptoms and the laboratory confirmation as observed in the patient line list (bars) and as estimated by the best-fitting negative binomial distribution (solid line represents the mean estimate; shaded area represents the 95%CI of estimates).*

We assessed possible bias in the data due to delays between the time of confirmation of a case and the time when the patient record is inserted in the centralized database (i.e., the reporting date). To this aim, we compared the epidemic curves by symptom onset and by confirmation date in successive updates of the patient line list (Figure S4). From these curves, we computed the delay by which each positive case and its symptom (when symptomatic) are reported in the current version of patient line list (Figure S5), and we used maximum likelihood to fit to each of them a gamma distribution. The parameters of the best-fitting distributions are reported in Tab. S2.

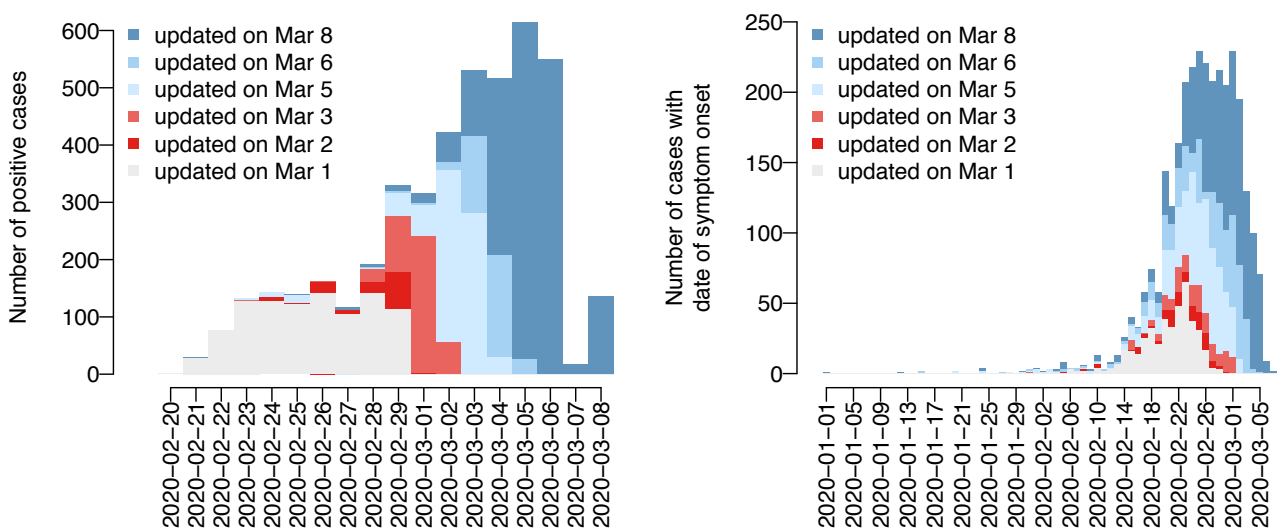

***Figure S4.*** *Left: Number of positive COVID-19 cases by date of laboratory test results. Right: Number of positive COVID-19 cases by date of symptom onset. Data are shown by date of inclusion in the patient line list.*

Figure S6 shows that both reporting delays declined over time suggesting a progressive improved reliability of the database as regards cases occurred soon before the update. Figure S7 shows the empirical cumulative distribution function of the delay from symptom onset to reporting and highlights the fact that there is a delay of about two weeks until we can assume 90% completeness of the data for the recorded cases with symptom onset at a given date.

***Table S2.*** *Parameters of the gamma distributions used for fitting the observed distributions of delay to reporting (from symptom onset and laboratory test results).*

| Delay to reporting | From symptom onset (mean and se) | From laboratory test results (mean and se) |
| --- | --- | --- |
| Shape (standard deviation) | 1.88 (0.055) | 2.13 (0.078) |
| Rate (standard deviation) | 0.26 (0.0087) | 0.59 (0.024) |
| Mean (95%CrI) | 7.3 (0.8-20.8) | 3.6 (0.5-9.9) |

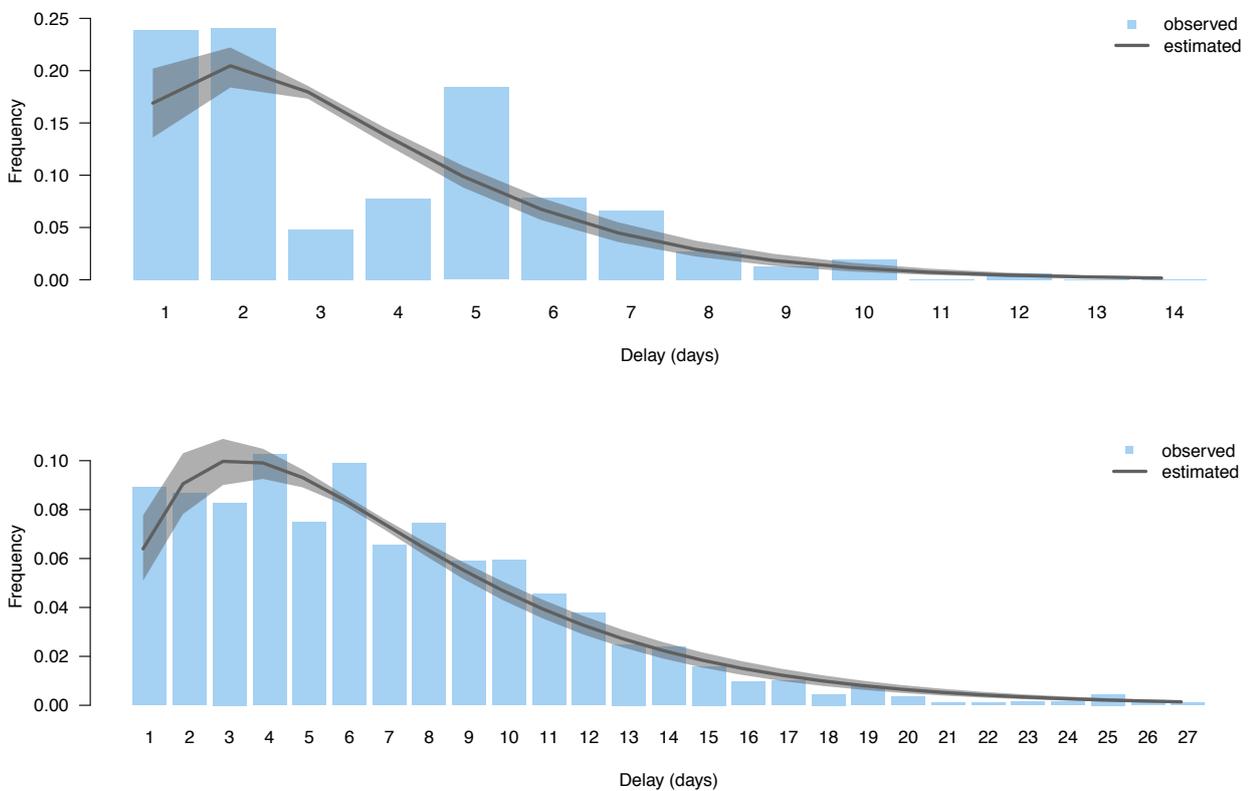

***Figure S5.*** *Top: Distribution of the reporting delays for the date of laboratory test results. The bars represent the data and solid line represents the mean of the best fitting gamma distribution (the shaded area represents 95%CI). Bottom: As Top, but for the delay for the symptom onset dates.*

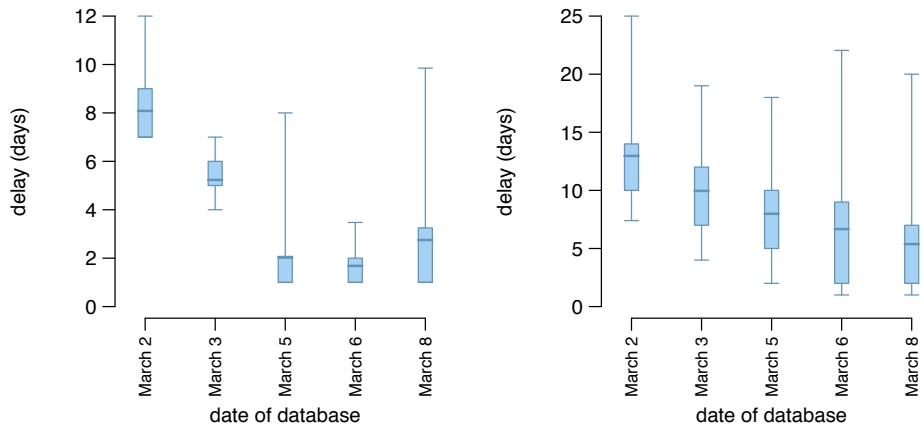

***Figure S6.*** *Distributions of reporting delays for successive updates of the database. Left: Delay from laboratory test results. Right: Delay from symptom onset.*

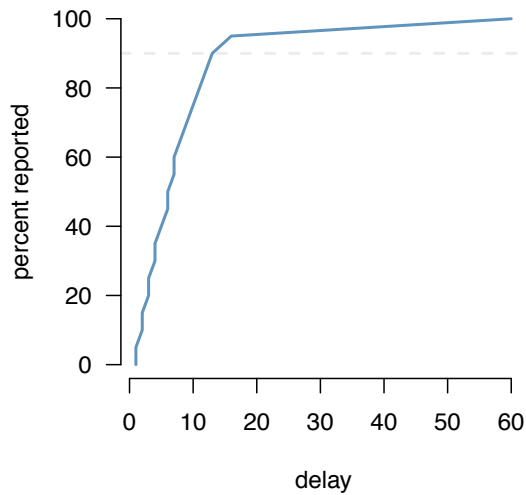

***Figure S7.*** *Empirical cumulative distribution function for the delays between symptom onset and reporting in the current patient line list.*

### 4. Serial interval distribution

Data on close contacts' of positive cases were collected through standardized interviews of performed by the ATS operators of the Lombardy Region. Epidemiological links were used to estimate the serial intervals for pairs of primary infectors and secondary cases between cases in the same cluster, connected by an epidemiological link.

Of the 6,106 reported contacts, only 3.8% resulted in COVID-19 infection. Of these 233, for 90 pairs of cases we found an infector-infectee relationship and have the dates of symptom onset of both cases. Identified secondary cases were ascribed to 55 unique primary cases.

A gamma distribution was fitted to the obtained set of individual serial intervals by using maximum likelihood estimation. The best fitting gamma distribution (Fig. S8) has mean 6.6 days and median 5.5 days (shape 1.87, SD 0.26; scale 0.28, SD 0.04).

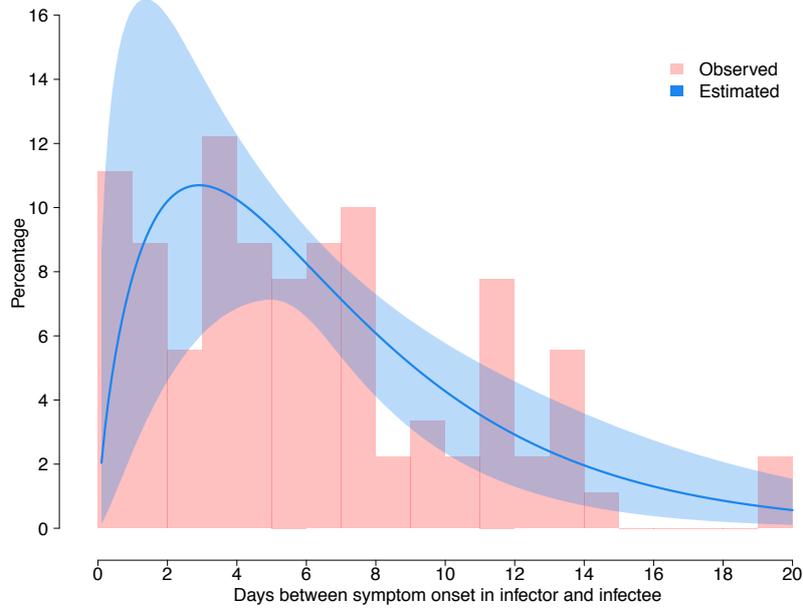

*Figure S8. Red bars represent observed lags between the symptom onset dates of secondary cases and their corresponding primary infectors. The solid line and the shaded area show the estimated mean and 95%CI of the serial interval as obtained by the best-fitting gamma distribution.*

### 5. Estimation of the reproduction number

The basic reproduction number $R_0$ represents the average number of secondary cases generated by a primary infector in a fully susceptible population. In general terms, when $R_0$ is larger than 1 the infection may spread in the population and the larger $R_0$ the larger effort required to control the epidemic. Once the number of susceptible individuals declines, the transmission potential of the disease at a given time t is measured in terms of the net reproduction number $R(t)$. The net reproduction number is useful to track the effectiveness of performed control measures and other factors affecting the spread of the epidemic (e.g., the behavioral response of the population) over time. As soon as $R(t)$ falls below 1, the epidemic starts to decline.

To estimate $R(t)$, we use the same methodology presented in reference [2-4]. We assumed that the daily number of new cases (date of symptom onset) with locally acquired infection $L(t)$ can be approximated by a Poisson distribution according to the equation

$$L(t) \sim \text{Pois}\left(R(t) \sum_{s=0}^{t} \varphi(s) C(t-s)\right)$$

where
- $C(t)$, with t from 1 to T, is daily number of new cases (date of symptom onset);
- $R(t)$ is the net reproduction number at time t;
- $\varphi(s)$ is the distribution of the generation time (corresponding to the distribution of the serial interval) calculated at time s.

The likelihood $\mathcal{L}$ of the observed time series of cases from day 1 to day T conditional on $C(0)$ is thus given by

$$\mathcal{L} = \prod_{t=1}^{T} P\left(L(t); R(t) \sum_{s=1}^{t} \varphi(s) C(t-s))\right)$$

where P(k; λ) is the probability mass function of a Poisson distribution (i.e., the probability of observing k events if these events occur with rate λ).

We then used MCMC Metropolis-Hastings sampling to estimate the posterior distribution of R(t). To estimate $R_0$, we assumed that during the period where the epidemic showed exponential growth R(t)=$R_0$ and used the above described procedure.

## 6. Sensitivity analysis on the serial interval

As sensitivity analysis, we report the results obtained by considering estimates of the serial interval obtained by epidemiological investigations of the very first few clusters in Wuhan, China [5] and by the analysis of transmission the Chinese provinces outside Hubei [4]. The estimated trends in the dynamics of R(t) are robust with these choices of the serial interval (Fig. S9 and S10). In particular, the decreasing trend of R(t) since late February is clearly visible in both scenarios. Clearly, in absolute terms, the longer the serial interval and the larger is the estimated R(t) value.

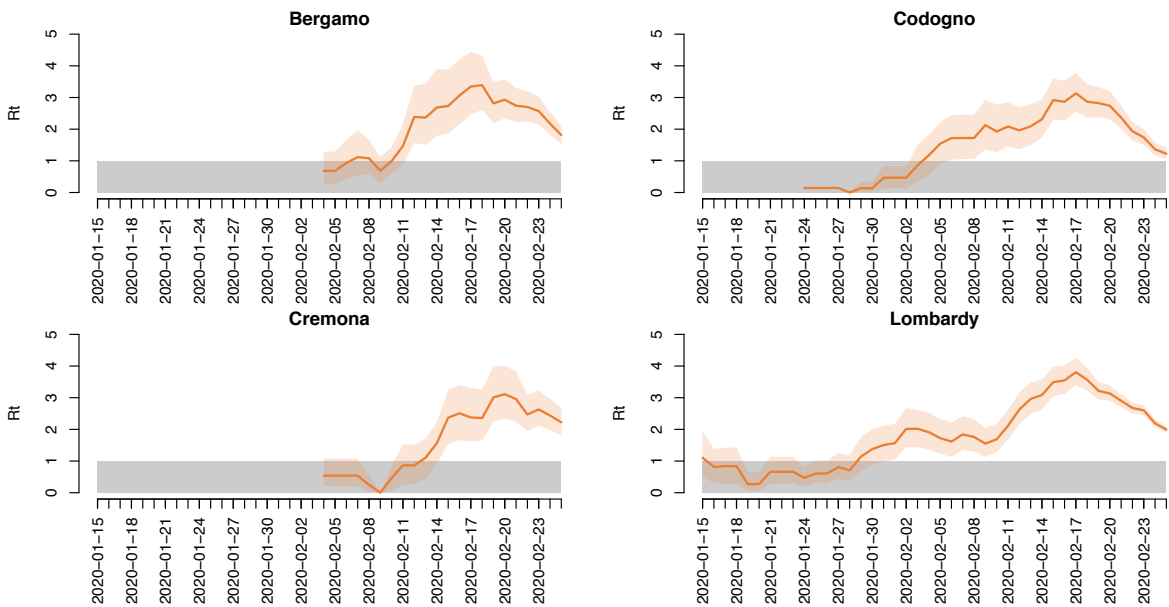

*Figure S9.* *Estimated net reproduction number (Rt) in the three clusters and in Lombardy as a whole, over a 4-day moving average by considering a serial interval of mean 5.1 days (SD: 3.4 days) [4]. We excluded cases with symptom onset after February 25, 2020 to account for the incompleteness of the dataset in the last few days due to reporting delays.*

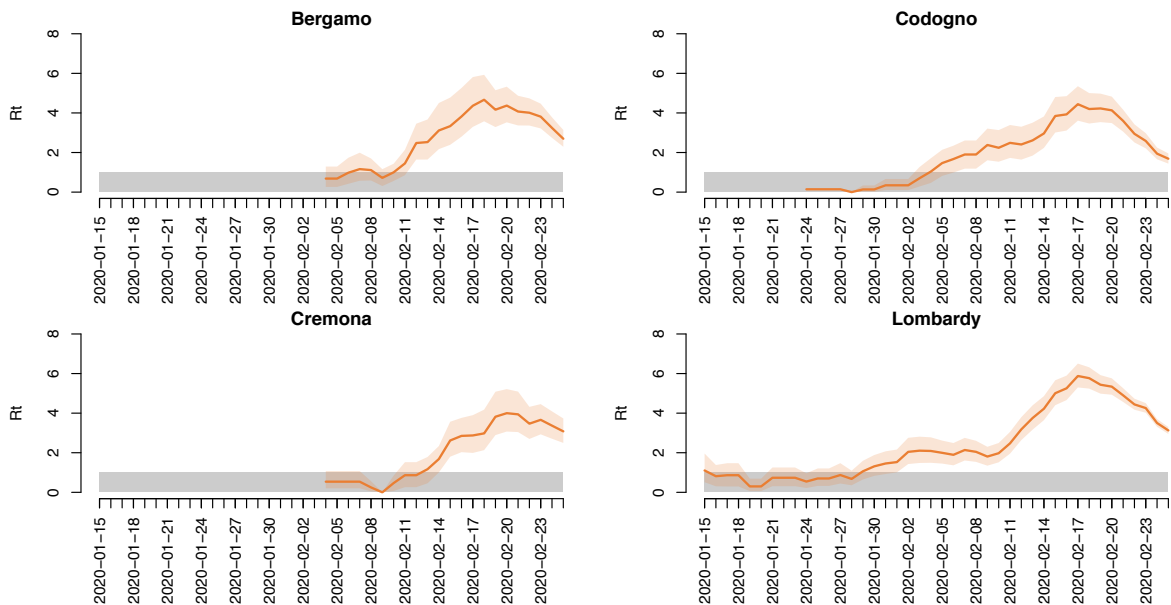

***Figure S10.*** *Estimated net reproduction number R(t) in the three clusters and in Lombardy as a whole, over a 4-day moving average by considering a serial interval of mean 7.5 days (SD: 3.4 days) [5]. We excluded cases with symptom onset after February 25, 2020 to account for the incompleteness of the dataset in the last few days due to reporting delays.*

## 7. References

1. The Italian National Institute of Statistics (ISTAT). Resident population in Italy in 2019. Available from: http://demo.istat.it/
2. World Health Organization Ebola Response Team. Ebola virus disease in West Africa--the first 9 months of the epidemic and forward projections. N Engl J Med 2014; 371(16): 1481-95.
3. Liu Q-H, et al. Measurability of the epidemic reproduction number in data-driven contact networks. Proc Natl Acad Sci 2018; 115(50): 12680.
4. Zhang J, et al. Evolving epidemiology of novel coronavirus diseases 2019 and possible interruption of local transmission outside Hubei Province in China: a descriptive and modeling study. medRxiv 2020; doi: https://doi.org/10.1101/2020.02.21.20026328
5. Li Q, et al. Early transmission dynamics in Wuhan, China, of novel coronavirus–infected pneumonia. N Engl J of Med 2020.

**+ LOMBARDY TASK FORCE CORONAVIRUS** Cajazzo L, Salmoiraghi M, Andreassi A, Sabatino G, Cornaggia N, Bodina A, Toso C, Crottogini L, Preziosi G, De Filippis G, Viganò P, Rizzardini G, Bergamaschi W, Ciconali G, Senatore S, Lamberti A, Perotti GM, Bianchi D, Bernocchi P, Bocconi A, Fugazza L, Storti E, Bisagni P, Bracchi M, Paglia S, Baracco A, Ragazzetti A, Archi D, Gambarana E, Raimondi L, Beccarini V, Piazza M, Cerri MC, Papetti C, Maffezzini E, Zoncada A, Ceccomanicini S, Ferraresi A, Fornabaio C, Fumarola B, Milesi M, Zambolin G, Cammelli L, Fornaciari M, Pezzetti F, Maghini G, Ferrari D, Grandi A, Giorgi-Pierfranceschi M, Nardecchia A, De Gennaro F, Cavalli I, Rizzi N, Milanetti F, Toffolon F, Maninetti L, Dizioli P, Orizio P, Bosio G, Betti M, Maestrelli M, Stifani I, Bagliovo F, Martinelli M, Torres A, Viola E, Cuzzoli A, Beccara L, Coluccello A,

Machiavelli A, Poli N, Frittoli A, Rocca L, Bussi D, Lolli M, Barbieri C, Galli P, Barbarini M, Bernardelli A, Gianotti G, Vismarra M, Nicora C, Triarico A, Marena C, Muzzi A, Mojoli F, Perlini S, Palo A, Barbarini D, Bruno A, Cambieri P, Campanini G, Comolli G, Corbella M, Daturi R, Furione M, Mariani B, Maserati R, Monzillo E, Paolucci S, Parea M, Percivalle E, Piralla A, Sarasini A, Zavattoni M, Adzasehoun G, Bellotti L, Cabano E, Casali G, Dossena L, Frisco G, Garbagnoli G, Girello A, Landini V, Lucchelli C, Maliardi V, Pezzaia S, Premoli M, Bonetti A, Caneva G, Cassaniti I, Corcione A, Di Martino R, Di Napoli A, Ferrari A, Ferrari G, Fiorina L, Giardina F, Mercato A, Novazzi F, Ratano G, Rossi B, Sciabica IM, Tallarita M, Vecchio NE, Binda S, Galli C, Auxilia F, Anselmi G, Primache V, Pellegata G, Sfogliarini R, Sala PM, Ficarelli M, Viganò G, Merli G, Canetta C, Scartabellati A, Buscarini E, Lapiana G, SinatraML, Bona A, Mantoan C, Giroletti A, Passera S, Frati P, Tacca MR, Griffanti P, Stasi MB, Pezzoli F, Fumagalli MA, Limonta F, Cacciabue E, Bombardieri G, Canini S, Scetti S, Pagani G, Picciché A, Cannistraro V, Frattini S, Rizzi M, Fagiuoli S, Di Marco F, Cosentini R, Lorini LF, Colledan M, Cesa S, Zanotta A, Rota L, Ferrari M, Capelli C, Tomasoni L, Ferrari T, Gatti V, Gregis F, Gamba E, Vitalini MG, Vezzosi L, Piro A, Rossi C, Basili D, Anghinoni E, Rubagotti L, Troiano G, Firmi A, Bocchi M, Pirali F, Finardi C, Nolli F, Azzi M, Silva S, Cadum E, Ancarani C, Lodola S, Boccardi C, Bai S, Edo S, Riboli S, Della Valle GP, Sileo C, Lanfredini L, Fagandini F, Marazza G, Caruana A, Gasparotti C, Campa I, Zanella A, Pesenti A, Tagliabue P, Zangrillo A